%% file: Main_Manuscript_ISBI.tex
\documentclass{article}

\usepackage[final]{neurips_2024}
    
\usepackage[utf8]{inputenc} 
\usepackage[T1]{fontenc}    
\usepackage{hyperref}       
\usepackage{url}            
\usepackage{booktabs}       
\usepackage{amsfonts,amsmath} 
\usepackage{nicefrac}       
\usepackage{microtype}      
\usepackage{lipsum}		
\usepackage{graphicx}
\usepackage{natbib}
\usepackage{doi}

\input{preamble}

\input{macros}

\title{End-to-End Model-based Deep Learning for Dual-Energy Computed Tomography Material Decomposition}

\author{\hspace{-.3cm} Jiandong~Wang$^{1, 2}$ \hspace{3.5cm} Alessandro~Perelli$^2$\\
\hspace{.2cm} $^1$Shenzhen Xilaiheng Medical Electronics \hspace{.4cm} $^2$Centre for Medical Engineering and Technology\\
\hspace{.6cm} (HORRON), China \hspace{2.6cm} University of Dundee, DD1 4HN, UK\\
\hspace{.1cm} \texttt{jack@horron.com} \hspace{3cm} \texttt{aperelli001@dundee.ac.uk}
}

\hypersetup{
pdftitle={End-to-End Model-based Deep Learning for Dual-Energy Computed Tomography Material Decomposition},
pdfsubject={Medical Imaging, Deep Learning},
pdfauthor={Jiandong~Wang, Alessandro~Perelli},
pdfkeywords={Dual Energy Computed Tomography, Material decomposition, Deep learning, Optimization},
}

\begin{document}
\maketitle

\begin{abstract}
    Dual energy X-ray Computed Tomography (DECT) enables to automatically decompose materials in clinical images without the manual segmentation using the dependency of the X-ray linear attenuation with energy. In this work we propose a deep learning procedure called End-to-End Material Decomposition (E2E-DEcomp) for quantitative material decomposition which directly convert the CT projection data into material images. The algorithm is based on incorporating the knowledge of the spectral model DECT system into the deep learning training loss and combining a data-learned prior in the material image domain. Furthermore, the training does not require any energy-based images in the dataset but rather only sinogram and material images. We show the effectiveness of the proposed direct E2E-DEcomp method on the AAPM spectral CT dataset \citet{sidky2023report} compared with state of the art supervised deep learning networks. 
\end{abstract}

\input{intro}

\input{DECT}

\input{Decomp-MoDL}

\input{Results}

\bibliographystyle{plainnat}
\bibliography{references}  

\end{document}

%% file: preamble.tex
\usepackage{amsmath, amssymb, amsfonts}
\usepackage{algorithm, algcompatible}
\usepackage{url}
\usepackage{array}
\usepackage{graphicx}

\usepackage{caption}
\usepackage{subcaption}

\usepackage{adjustbox}
\usepackage{textcomp}
\usepackage{verbatim}
\usepackage{multirow}

\usepackage{cite}
\usepackage{color}
\usepackage{bm}     

\makeatletter
\def\hlinewd#1{%
	\noalign{\ifnum0=`}\fi\hrule \@height #1 \futurelet
	\reserved@a\@xhline}
\makeatother

\newcolumntype{?}{!{\vrule width 1.5pt}}

\usepackage{pgfplots}
\usepackage{pgf}
\usepackage{tikz}
\usetikzlibrary{matrix}
\usetikzlibrary{decorations.pathmorphing} 
\usetikzlibrary{fit}					  
\usetikzlibrary{backgrounds}	          
\usetikzlibrary{shapes.arrows}
\usetikzlibrary{shapes.geometric}
\usetikzlibrary{shapes.multipart}
\usetikzlibrary{matrix, arrows, decorations.pathreplacing, positioning}
\tikzstyle{rec}=[draw,rectangle, minimum height=2cm]
\tikzset{>=stealth', punkt/.style={rectangle, 
		fill=gray!40, 
		draw=black, very thick, text width=3em, minimum height=2.5em, text centered}}
\tikzset{>=stealth', Denoi/.style={rectangle, fill=blue!20, 
		draw=black, very thick, text width=6em, minimum height=3.5em, text centered}}
\tikzset{>=stealth', CG/.style={rectangle, 
		fill=green!20, draw=black, very thick, text width=9.5em, minimum height=3.5em, text centered}}
\tikzstyle{background} = [rectangle, fill=green!20, inner sep=0.1cm, rounded corners=4mm, 
                          minimum height=6em]
\tikzstyle{sum}   = [draw, fill=gray!40, circle, node distance=1cm]
\tikzstyle{dot}   = [circle, fill=black, inner sep=0pt, minimum size=5pt, node contents={}]
\tikzstyle{fig_n} = [node distance=30pt, inner sep=0cm]


%% file: macros.tex
\def\*#1{\mathbf{#1}}
\def\+#1{\mathcal{#1}}
\def\-#1{\mathbb{#1}}
\def\~#1{\mathrm{#1}}
\def\R{\mathbb{R}}

\DeclareMathOperator{\diag}{diag}

\algnewcommand\INPUT{\item[\textbf{Input:}]}
\algnewcommand\OUTPUT{\item[\textbf{Output:}]}

\newcommand{\boldr}{\bm{r}}

\newcommand{\boldmu}{\bm{\mu}}

\newcommand{\boldA}{\bm{A}}

\newcommand{\calL}{\mathcal{L}}

\newcommand{\rmd}{\mathrm{d}}

\renewcommand{\hbar}{\bar{h}}

\newcommand{\argmin}{\operatornamewithlimits{arg\,min}}

\newcommand{\transp}{^\top}

%% file: intro.tex
\section{Introduction}\label{sec:introduction}

Dual-energy computed tomography (DECT) is one spectral CT technology which is based on the deployment of two X-ray sources at different energies which can potentially allow to discriminate different materials in a specimen or to reconstruct virtual mono-energetic images which is of utmost interest in clinical imaging applications and industrial non-destructive testing \citet{mendoncca2013flexible}. The dependency of the attenuation coefficient of different materials respect to the X-ray energy can be leveraged in the DECT material decomposition procedure whose aim is to estimate each pixel's value as a linear combination of two different basis materials \citet{johnson2007material}. 

Different approaches have been developed to obtain material images: the image domain techniques are based on first reconstructing independently the energy dependent attenuation in each pixel \citet{maass2009image}. The majority of the proposed networks such as U-Net \citet{nadkarni2022material} and generative adversarial network (GAN) \citet{shi2021material} are trained with a supervised learning approach which requires the pair of energy reconstructed Dual-energy computed tomography (DECT) images and basis material segmented images in the dataset. However these methods do not account from the beam-hardening effect, caused by the poly-energetic nature of the X-ray source. Moreover this approach leads to propagate the estimation errors from the reconstruction to the subsequent material decomposition. 

In order to account for the beam-hardening effect, one-step methods directly estimate basis materials images from \citet{cai2013full} measurements projects by leveraging a model-based optimization function to minimize. However because of the highly non-linear due to the energies X-ray source coupling, this leads to minimize non-convex cost functions which require high computational cost \citet{long2014multi}. 

An alternative approach is based on decomposing the high and low energy sinogram into two independent measurements which correspond to a single basis material. Different approximations of the decomposition function that convert the dual energy sinograms into  materials independent sinograms have been proposed in \citet{alvarez1976energy}. Afterwards each material sinogram is converted into the image domain using model-based optimization methods \citet{mechlem2018spectral} with spatial regularization. 

Recently, other works have exploited the paradigm of combining deep learning and the knowledge of the physics of the DECT model within the optimisation problem. The one-step material decomposition is implemented using supervised unrolling algorithms in \citet{eguizabal2022a, perelli2021} or using the Noise2Inverse framework which uses pair of sub-sampled noisy sinograms and training dataset \citet{fang2021}. One limitation of these methods is the computational cost of the iterative solver which hinders the usage in applications which require strict time constraints. 

In this work, we propose a new optimization framework called End-to-End Material Decomposition (E2E-DEcomp) based on the idea to directly embedding the material decomposition function into the model-based optimization function. The optimization problem to solve becomes linear and the mapping is learned from the data during the iteration procedure of the solver.

Furthermore, the proposed method does not require any system calibration procedures to determine the material decomposition function as needed in previous approaches. The designed cost function contains the data consistency term in the material sinogram domain and a regularization term, acting in the material image domain, which is learned through an implicit denoising neural network-based function.

%% file: DECT.tex
\section{Dual-Energy CT Forward Model}

We introduce the forward mathematical model of the DECT system where two sources emit poly-energetic X-ray photons with spectrum $S_k(E)$, where $k\in\{e_1,e_2\}$, and the array of integrating detectors with dimension $N$ collects the photons after attenuation during the two independent acquisitions.   

The linear attenuation is a spatially- and energy-dependent function $\mu(\boldr,E)$ at position $\boldr\in\R^n$ and energy  $E\in \R_+$. The energy-dependent image $\mu$ is sampled on a grid with $M$ basis pixel-functions $u_m$ such that $\mu(\boldr,E) = \sum_{m=1}^M \mu_m(E) u_m(\boldr)$ where $\mu_m(E)$ is the energy-dependent attenuation at pixel $m$. The photons' intensities $I_{n,k}$ detected for the $n$-th ray and energy source spectrum $k$, are a sum of Poisson random variables. The conditional mean of $I_{n,k}$ is 
\begin{equation}\label{eq:cond_count}
	\-E\left[I_{n,k} | \bm\mu\right] = \int_{0}^{\infty} S_k(E) e^{-\sum_{m=1}^M A_{n,m} \mu_m(E)} dE
\end{equation}
with $\boldA \in \R^{N\times M}$ defined as $[\boldA]_{n,m} = \int_{\calL_n} u_m(\boldr) \,\rmd \boldr$ and $\boldmu(E) = [\mu_1(E),\dots,\mu_J(E)]\transp \in \R_+^M$ is the discretized energy-dependent attenuation. In the normal dose case, the DECT measurements $\*y\in\-R^{N\times 2}$ collected from the scanner are obtained using the negative logarithm of the measured photons at each energy $\*y_n = [y_{n,e_1}, y_{n,e_2}] = \left[  -\log\left( I_{n,e_1} \right), -\log\left( I_{n,e_2} \right)\right]$. The conditional mean of the measurements over the linear attenuation becomes
\begin{equation}\label{eq:log_avg}
	\-E\left[\*y_n | \bm\mu\right] = -\log\left(\int_{0}^{\infty} \*S(E) e^{-\sum_{m=1}^M A_{n,m} \mu_m(E)} dE\right)
\end{equation}
with $\*S(E) = \left[ S_{e_1}(E), S_{e_2}(E) \right]$ representing the spectrum of the two X-ray sources $\{e_1, e_2\}$ for each energy $E$. 

From the model \eqref{eq:log_avg} we can derive the conditional mean $\-E\left[\*y_n | \*x\right]$ over the material vectorized images $\*x\in\-R^{M\times 2}$, by exploiting the decomposition of the attenuation as a linear combination of the mass density of two basis materials
\begin{equation}\label{eq:m_decomp}
	\mu_i(E) = x_{m,1}\varphi_1(E) + x_{m,2}\varphi_2(E)
\end{equation}
where $x_{m,s}$ (mg/cm$^3$) is the equivalent density for basis materials $s$ at voxel $m$ and $\varphi_s(E)$ (cm$^2$/mg) is the known energy-dependent mass attenuation function for basis material $s$. Then, by substituting Eq.~\eqref{eq:m_decomp} into Eq.~\eqref{eq:log_avg}, we have
\begin{equation}\label{eq:log_avg_mat}
	h(\*p_n) = -\log\left(\int_{0}^{\infty} \*S(E) e^{-\*p_n\left(\bm\varphi(E)\right)^T} dE\right)
\end{equation}
where $\bm\varphi(E) = \left[\varphi_1(E), \varphi_2(E)\right]$, $\*p_n$ (mg/cm$^2$) is the material density projection defined as
\begin{equation}
	\*p_n\left(\bm\varphi(E)\right) = \sum_{m=1}^M A_{n,m} \left(x_{m,1}\varphi_1(E) + x_{m,2} \varphi_2(E) \right) 
\end{equation}
and $h: \-R^2 \rightarrow \-R^2$ is a vector-valued function which models the non-linear relationship between the material density projections and the expected attenuation. From this, we have $\-E\left[\*y_n |\*x \right] = h\left([\*A\*x]_n\right)$. The inverse function $h^{-1}: \-R^2 \rightarrow \-R^2$ is defined as $h^{-1}\left(h(\*p_n) \right) = \*p_n$ which represents the material decomposition in the sinogram domain since it converts the energy sinogram $h(\*p_n)$ into the material sinograms $\*p_n$.

\section{Model-based Optimization Problem}

In the normal X-ray dose case, we can approximate the Poisson distribution of the measurements $I_{n,k}$ with an anisotropic Gaussian distribution over $\*y_n$ with conditional average $\-E\left[\*y_n |\*x \right]$ and diagonal covariance $\*W$, where individual projections are conditional independent. The negative log-likelihood (NLL) $f(\*y, \*x) = -\log P(\*y | \*x)$ is expressed using the expression of the conditional mean \eqref{eq:log_avg_mat} as follow
\begin{equation}\label{eq:NLL_h}
f(\*y, \*x) = \frac{1}{2} \sum_{n=1}^N\| \*y_n - h\left([\*A\*x]_n\right) \|^2_{\*W_n} + C
\end{equation}
where $C$ is a normalizing constant, and $\*W_n = \~{Cov}^{-1}(\*y_n | \*x) = \diag([ w_{n,e_1}, w_{n,e_2} ])$ is the inverse covariance of $\*y_n$ with $w_{n,k} = [\~{Var}(\*y_{n,k} | \*x)]^{-1} \approx I_{n,k}, \; k \in \{ e_1, e_2 \}$.

Model-based iterative reconstruction (MBIR) approaches aim at optimising the cost function \eqref{eq:NLL_h}, however this problem is computationally demanding to solve because $f(\*y, \*x)$ is non-linear and the inverse function $h^{-1}$ is challenging to compute since it requires an accurate calibration procedure.

%% file: Decomp-MoDL.tex
\section{End-to-End Model-based Deep Learning for Material Decomposition (E2E-Decomp)}

We describe the idea of embedding a learned material decomposition polynomial network $P_{\theta}$, whose parameters $\theta$ are learned from the sinogram data, into the negative log-likelihood (NLL) term. Using the definition $h^{-1}\left(h([\*A\*x]_n) \right) = [\*A\*x]_n
$ and the DECT model \eqref{eq:log_avg_mat}, we obtain the estimated material sinogram
\begin{equation}
    \hat{\*p}_n = h^{-1}(\*y)_n \approx \+P_{\theta}(\*y)_n 
\end{equation}
where the last approximation is obtained by substituting the vector function $\+P_{\bm\theta} \colon \R^{N\times 2}_+ \to \R^{N\times 2}_+$ with learned parameters $\bm\theta$ for the non-linear material decomposition function $h^{-1}$. $\+P_{\theta}$ is a vector polynomial function which takes as input the energy sinogram $\*y\in\R^{N\times 2}$ and output the material sinogram of the same dimension as follow
\begin{equation*}
	\+P_{\theta}(\*y)_n =[\hat{p}_{1,n}, \hat{p}_{2,n}] = \left[
	\begin{array}{c}
		\sum_{i=0}^I \sum_{j=0}^J \theta_{i,j,e_1} y_{n,e_1}^i y_{n,e_2}^j \\
		\sum_{i=0}^I \sum_{j=0}^J \theta_{i,j,e_2} y_{n,e_1} y_{n,e_2} \\
	\end{array}
	\right]
\end{equation*}
where the matrix of parameters $\bm\theta$ is trained through a polynomial regression network. The NLL function, which guarantees consistency in the material sinogram domain, is 
\begin{equation}\label{eq:data_cons}
    f(\*y, \*x) = \sum_{n=1}^{N} \|\+P_{\bm\theta}(\*y)_n - [\*A \*x]_n \|^2_{\*B_n}
\end{equation}
where $\*B_n = [\nabla \+P_{\bm\theta}(\*y)_n]^{-1}\, \*W_n\, [\nabla \+P_{\bm\theta}(\*y)_n]^{-T}\in \R^{2\times 2}$ is the statistical covariance matrix of the problem in the new domain which is obtained using the first order Taylor expansion as in \citet{zhang2013model} using the polynomial network $\+P_{\theta}$, where $\nabla \+P_{\bm\theta}(\cdot)$ can be efficiently computed using back-propagation. 

The material decomposition of the images $\*m\in\R^{M\times 2}$ is formulated as a minimization of the following cost function $\*x^* = \argmin_{\*x\in\R^{M\times 2}_+} f(\*y,\*x) + \lambda R(\*x)$ where $f(\*y,\*x)$ represents the DECT data consistency based on the mathematical negative log-likelihood model in \eqref{eq:data_cons}, $\lambda$ is the regularization parameter and the regularization $R(\*x) = \|\+R_{\bm\rho}(\*x)\|^2$ is learned from the materials' data. We constrained the optimisation to non-negative material images. We design $\+R_{\bm\rho}$ to be a learned convolutional neural network (CNN) estimator of noise of the material images, with parameters $\bm\rho$
\begin{equation}\label{eq:reg_r}
	\+R_{\bm\rho}(\*x) = (\*I - \+D_{\bm\rho})(\*x) = \*x - \+D_{\bm\rho}(\*x)
\end{equation}
where $\+D_{\bm\rho}(\*x)$ is the denoised version of $\*x$, after the removal of noise. By formulating the problem with \eqref{eq:data_cons}-\eqref{eq:reg_r}, we are able to decouple the learning procedures which are applied in different domain since the data consistent term contains the learning module $\+P_{\bm\theta}(\cdot)$ for the material decomposition in the sinogram domain while the learning of the image data prior $\+R_{\bm\rho}(\cdot)$ is applied within the regularization term as
\begin{equation}\label{eq:opt_fully}
	\*x^* = \arg\min_{\*x\in\R^{M\times 2}_+} \sum_{n=1}^{N} \|\+P_{\bm\theta}(\*y)_n - [\*A \*x]_n \|^2_{\*B_n} +\lambda \| \*x - \+D_{\bm\rho}(\*x) \|^2
\end{equation}

Solving \eqref{eq:opt_fully} with any first order solvers would require computing the Jacobian $\*J$ of the denoiser $\+D_{\bm\rho}(\cdot)$ which is computationally demanding. We adopt the same approach as in \citet{aggarwal2018modl} by approximating the non-linear term $\+D_{\bm\rho}(\*x)$ using the first order Taylor approximation at the $k$-th iteration which leads to $\| \*x - \+D_{\bm\rho}(\*x^k + \nabla\*x) \|^2 \approx \| \*x - \*z^k \|^2 + \| \*J^k\Delta\*x \|^2$. For small perturbations around $\*m^k$, the term $\| \*J^k\Delta\*x  \|^2$ can be approximated to zero. Therefore the problem \eqref{eq:opt_fully} becomes
\begin{eqnarray}\label{eq:opt_split}
    \*x^k &=& \argmin_{\*x\in\R^{M\times 2}_+} \sum_{n=1}^{N} \|\+P_{\bm\theta}(\*y)_n - [\*A \*x]_n \|^2_{\*B_n} +\lambda \| \*x - \*z^k \|^2 \nonumber \\
    \*z^{k+1} &=& \+D_{\bm\rho}(\*x^k)
\end{eqnarray}

\noindent The optimization problem in the first step of problem \eqref{eq:opt_split} has a quadratic form therefore the associated linear problem 
\begin{equation}\label{eq:data_consist}
    \left[\*A^T(\*B\odot \*A) +\lambda\*I\right]\*x^k = \left( \*A^T \+P_{\bm\theta}(\*y) + \lambda \*z^k \right)
\end{equation}
where $\*B=\left[ \diag(\*B_1) ; \ldots ; \diag(\*B_N)\right]\in\R^{N\times 2}$ is the concatenation of the diagonal element of the covariance weighting matrices $\*B_n$. Eq. \eqref{eq:data_consist} can be solved efficiently using using the Conjugate Gradient (CG) iterative algorithm \citet{nazareth2009conjugate} which does not require any computation of the matrix inverse. 

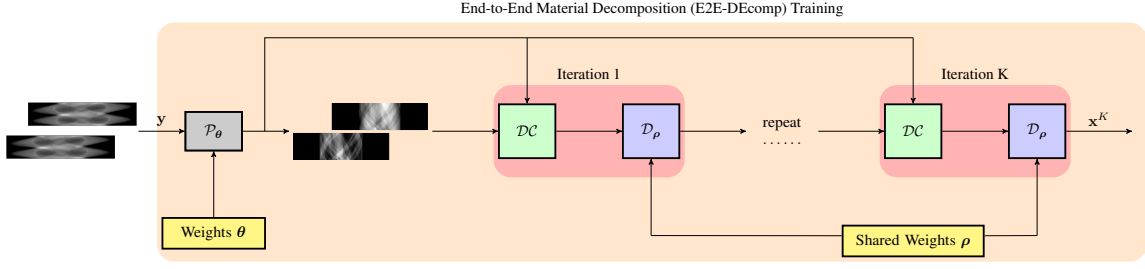
\begin{figure*}[!t]
	\centering
	\resizebox{\textwidth}{!}{
		\begin{tikzpicture}[node distance=1.5cm,scale=1]
			\node [punkt] (P) {$\+P_{\bm\theta}$};
			\node[fig_n, left= of P, inner sep=0pt] (IN) 
			{\includegraphics[width=.2\textwidth]{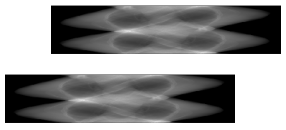}};	
			\path[->] (IN.east) edge node [above] {$\*y$} (P.west);
			\node [punkt] (P) {$\+P_{\bm\theta}$};
			\node[fig_n, right= of P, inner sep=3pt] (DEC) 
			{\includegraphics[width=.2\textwidth]{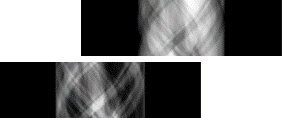}};	
			\node [CG, right= of DEC, text width=3em] (DC) {$\+{DC}$};		
			\path[->] (P.east) edge node {} (DEC.west);
			\node [Denoi, right= of DC, text width=3em] (Denoi) {$\+D_{\bm\rho}$};
			\path[->] (DC.east) edge node {} (Denoi.west);		
			\node [below= of Denoi, text width=0em, text centered] (EMPTY)  {};
			\path[->] (EMPTY.south) edge node {} (Denoi.south);
			\path[->] (DEC.east) edge node {} (DC.west);
			
			\node[right= of Denoi, text width=4em, text centered] (Iter) {repeat \\ $\ldots\ldots$};
			
			\node [CG, right= of Iter, text width=3em, text centered] (DC1) {$\+{DC}$};		
			\path[->] (Denoi.east) edge node {} (Iter.west);
			\path[->] (Iter.east) edge node {} (DC1.west);
			\node [Denoi, right= of DC1, text width=3em] (Denoi1) {$\+D_{\bm\rho}$};
			\path[->] (DC1.east) edge node {} (Denoi1.west);
			\node [right= of Denoi1, text width=0em, text centered] (OUT) {};
			\path[->] (Denoi1.east) edge node [above] {$\*x^K$} (OUT.west);
			\node [below= of Denoi1, text width=0em, text centered] (EMPTY1)  {};
			\path[->] (EMPTY1.south) edge node {} (Denoi1.south);
			\node [right= of Denoi1, text width=0em, text centered] (OUT) {};
			\node [Denoi, below= of DC1, text width=8.5em, minimum height=2em, fill=yellow!60] (W)  {Shared Weights $\bm\rho$};
			\path[-] (W.east)+(0, 0.12) edge node {} (EMPTY1.south);
			\path[-] (W.west)+(0, 0.12) edge node {} (EMPTY.south);
			\node [Denoi, below= of P, minimum height=2em, fill=yellow!60] (theta)  {Weights $\bm\theta$};
			\path[->] (theta.north) edge node {} (P.south);
			\node [above= of DC, text width=0em, text centered] (EMPTY2)  {};
			\path[->] (EMPTY2.south) edge node {} (DC.north);
			\node [above= of DC1, text width=0em, text centered] (EMPTY3)  {};
			\path[->] (EMPTY3.south) edge node {} (DC1.north);
			\path[-] (EMPTY2.south) edge node {} (EMPTY3.south);
			\draw [-] (P.east)+(0.5,0) |- (EMPTY2.south);
			\begin{pgfonlayer}{background}[node distance=1cm]
				\node [background,fit=(P) (DC) (DC1) (Denoi) (Denoi1) (W) (theta) (EMPTY2) (OUT), fill=orange!20, label=above:End-to-End Material Decomposition (E2E-DEcomp) Training] (end_to_end) {};
				\node [background,fit=(DC) (Denoi), fill=red!30, label=above:Iteration 1] (iter1) {};
				\node [background,fit=(DC1) (Denoi1), fill=red!30, label=above:Iteration K] (iter1) {};
			\end{pgfonlayer}
	\end{tikzpicture}}
	\caption{End-to-End Material Decomposition (E2E-DEcomp) material decomposition training workflow. The network is obtained by unfolding $K$ iterations of the algorithm and training end-to-end from energy sinograms $\*y$ to material images $\*x^K$. Each iteration is constituted by the data consistency block $\-{DC}$ and the denoising module $\-D_{\bm\rho}$ whose trainable parameters $\bm\rho$ are shared through the iterations.}\label{fig:DMDL_workflow}
\end{figure*}

The workflow of the E2E-DEcomp algorithm at inference is shown in Fig.~\ref{fig:DMDL_workflow}, and the structure of the E2E-DEcomp algorithm for inference is reported in Table \ref{table:DE_Modl}.

\begin{algorithm}[!h]
	\caption{End-to-End Material Decomposition (E2E-DEcomp) for Material Decomposition}
	\label{table:DE_Modl}
	\begin{algorithmic}[1]
		\REQUIRE energy sinogram $\*y$, parameters $\bm\rho$ and $\bm\theta$, number of iterations $K$
		\INPUT initial state $\*z^0 (= \*0) \in \R^{M\times 2}$, $\*x_c = \*A^T \+P_{\bm\theta}(\*y)$
		\OUTPUT $\*x^K$
		\FOR{$k = 0, \ldots, K-1$}
        \STATE $\*x^k \colon \left\{\begin{array}{l} \left[\*A^T(\*B\odot \*A) +\lambda\*I\right]\*x^k = \*A^T \+P_{\bm\theta}(\*y) + \lambda \*z^k \\ \mbox{solved using CG algorithm} \end{array}\right.$
        \STATE $\*z^{k+1} = \+D_{\bm\rho}(\*x^k)$ 
		\ENDFOR
	\end{algorithmic}
\end{algorithm}

%% file: Results.tex
\section{Numerical Results}

We have tested the E2E-DEcomp algorithm using the AAPM Dual-energy CT scans for a breast model containing adipose and fibroglandular tissues \citet{sidky2023report} and simulating the DECT sinogram with the ASTRA Toolbox \citet{van2016fast} using a parallel-beam geometry with a detector of $1024$ elements. We used $I_0=10^5$ X-ray photons with the spectrum from \citet{sidky2023report} and adding Poisson noise. The spatial resolution of the images is set to $512\times 512$. We deploy an end-to-end strategy to train the E2E-DEcomp algorithm where we fix a-priori the number of iteration $K=3$ and we train the overall unrolled iterations by calculating the mean square error (MSE) between the material estimates and the ground truth $\*x^*$. 

\begin{figure*}[!h]
	\centering	
	\begin{adjustbox}{minipage=\linewidth}
	\subcaptionbox{Ground truth image}{\includegraphics[width=0.1865\textwidth]{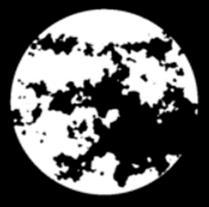}}
	\hspace{0.005\textwidth}
	\subcaptionbox{FBP $30$ angles}{\includegraphics[width=0.186\textwidth]{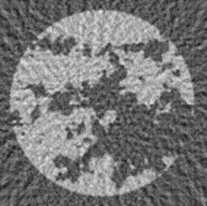}}
	\hspace{0.005\textwidth}
	\subcaptionbox{FBP $60$ angles}{\includegraphics[width=0.19\textwidth]{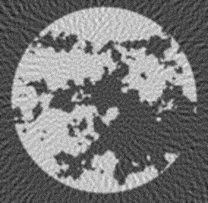}}
	\hspace{0.005\textwidth}
	\subcaptionbox{FBP $90$ angles}{\includegraphics[width=0.185\textwidth]{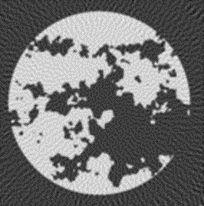}}
	\hspace{0.005\textwidth}
	\subcaptionbox{FBP $180$ angles}{\includegraphics[width=0.185\textwidth]{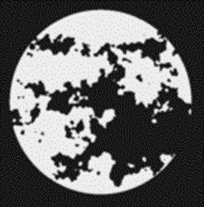}}
	\\
	\includegraphics[width=0.1865\textwidth]{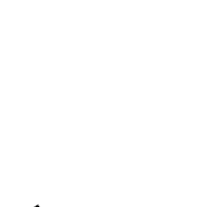}
	\hspace{0.005\textwidth}
	\subcaptionbox{E2E-DEcomp $30$ views}{\includegraphics[width=0.189\textwidth]{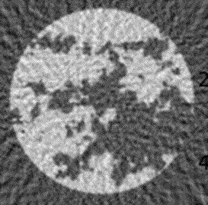}}
	\hspace{0.005\textwidth}
	\subcaptionbox{E2E-DEcomp $60$ angles}{\includegraphics[width=0.188\textwidth]{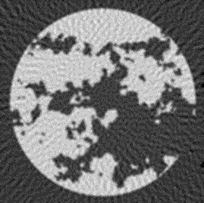}}
    \hspace{0.005\textwidth}
	\subcaptionbox{E2E-DEcomp $90$ angles}{\includegraphics[width=0.185\textwidth]{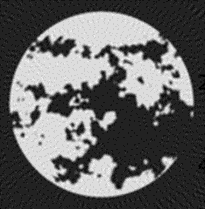}}
	\hspace{0.005\textwidth}
	\subcaptionbox{E2E-DEcomp $180$ angles}{\includegraphics[width=0.1865\textwidth]{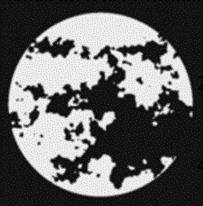}}
	\caption{Qualitative comparison between the material decomposition for adipose using E2E-DEcomp and FBP using different number of angular projections.}\label{fig:recon_spectral}
\end{adjustbox}
\end{figure*}

\begin{figure}[!h]
	\centering
	\begin{tikzpicture}[scale=1] 
		\begin{axis}[
			xlabel={Angles $\theta$},
			ylabel={PSNR (dB)},
			xtick={30, 90, 180, 360, 512},
			grid = major,
			legend columns=1,
			legend cell align=left,
			legend entries={E2E-DEcomp, FBP},
			legend style={at={(0.7,0.3)},anchor=north}
			]				
			\addplot[color=blue, mark=* , dashed] table[x=Ang, y=MoDL] {Noise_sim_MoDL_FBP.txt};
			\addplot[color=red, mark=square*, dashed] table[x=Ang, y=FBP] {Noise_sim_MoDL_FBP.txt};
			
		\end{axis}
	\end{tikzpicture}
	\caption{Comparison of DE-MoDL and FBP for 2 materials decomposition using noisy DECT acquisition with photon counts $I_0 = 10^5$. The PSNR metric is calculated for different number of DECT angular projections.} \label{fig:noise_sim}
\end{figure}
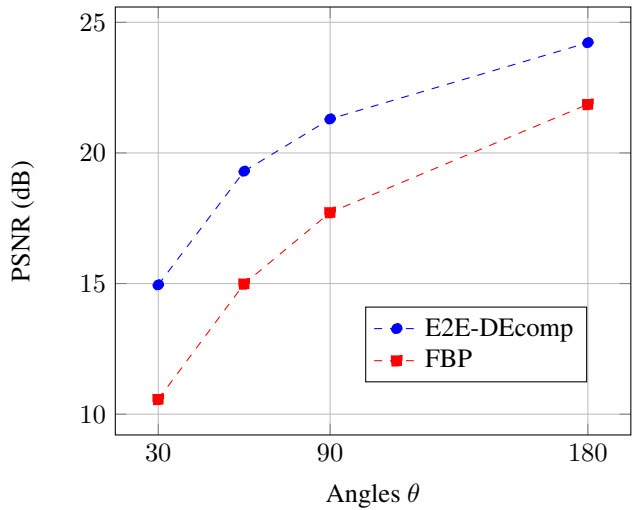

In order to reduce the number of learnable parameters we utilise the same architecture for the denoising module $\+D$ at each iteration $k$ with shared parameters $\bm\rho$. In Fig. \ref{fig:recon_spectral} it is shown the qualitative comparison on a test material image of the adipose tissue using filtered back projection (FBP) and E2E-DEcomp while in Fig. \ref{fig:noise_sim} is is reported the PSNR error for a set of 10 testing images for the 2 material decomposition.

\renewcommand{\arraystretch}{1.25}
\newcolumntype{C}{>{\centering\arraybackslash}p{0.07\textwidth}}

\begin{figure}[!h]
	\centering
	\begin{tikzpicture}[scale=1] 
		\begin{axis}[
			xlabel={Epochs},
			ylabel={PSNR (dB)},
			xtick={20, 40, 60, 80, 100, 120},
			grid = major,
			legend columns=1,
			legend cell align=left,
			legend entries={FBP ConvNet, E2E-DEcomp},
			legend style={at={(0.7,0.3)},anchor=north}
			]				
			\addplot[color=green, mark=* , dashed] table[x=Epochs, y=PSNR] {Training_DW.txt};
			\addplot[color=blue, mark=square*, dashed] table[x=Epochs, y=PSNR] {Training_DWDC.txt};
			
		\end{axis}
	\end{tikzpicture}
	\caption{Comparison of the PSNR training error between the FBP ConvNet and the E2E-DEcomp algorithms.} \label{fig:PSNR_training_DW}
\end{figure}
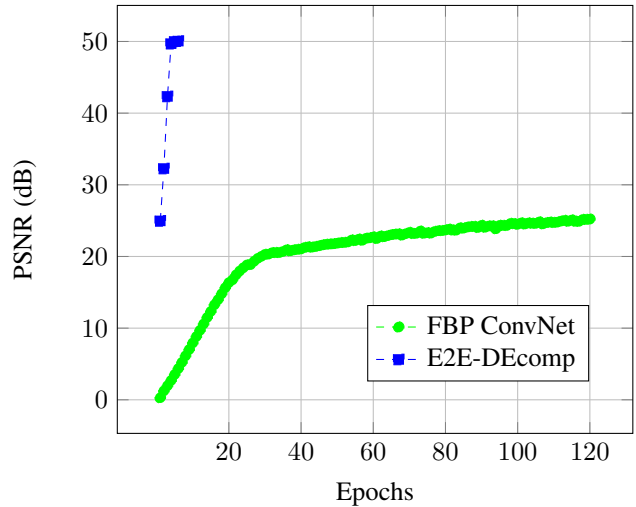

It is worth noting that the improvement in the decomposition accuracy are consistent, around 5 dB, across different levels of dose, i.e. from sparse views to higher number of projections. We have also compared the E2E-DEcomp framework with the FBP ConvNet method \citet{jin2017deep} and Fig. \ref{fig:PSNR_training_DW} shows how E2E-DEcomp can achieve a faster convergence in training using fewer epochs.

\section{Conclusion}

This work proposed a direct method for DECT material decomposition using a model-based optimization able to decouple the learning in the measurement and image domain. Numerical results show the effectiveness of the proposed E2E-DEcomp compared to other supervised approaches since it has fast convergence and excellent performance on low-dose DECT which can lead to further study with clinical dataset.

\section{Compliance with Ethical Standards}

This is a numerical simulation study for which no ethical approval was required.